\documentclass[9pt]{article}

\usepackage{tikz} 
\usepackage{amsmath}

\renewcommand{\mathbf}[1]{\ensuremath{\boldsymbol{#1}}}
\newcommand{\rmd}{\ensuremath{\mathrm{d}}}

\begin{document}

\title{The Machian contribution of the Universe to geodetic precession, frame dragging and gravitational clock effect }

\author {P. Christillin\\ \
Dipartimento di Fisica, \\
Universit\`a di Pisa\\
I.N.F.N. Sezione di Pisa\\ \\
and\\ \\
L. Barattini\\
Universit\`a di Pisa}

\maketitle

\begin{abstract}
Gravitomagnetism resulting from SR has been applied to geodetic precession and frame dragging.   
The substantial contribution of the ``fictitious'' Coriolis force, due to the relative rotation of the rest of the 
Universe in the non inertial frame of the free falling but rotating satellite, has to be taken into account, giving 
another quantitative confirmation of Mach's arguments and of the black hole nature of our Universe.  Also the 
gravitational clock effect has an elementary prediction in the present post Newtonian formulation.
\end{abstract}

\section{Introduction}

Recently a  set of ``Heaviside'' vector equations for gravity has been \emph{derived} from special relativity and 
shown to predict in simple terms the quadrupole gravitational radiation \cite{christbara}.

They are effective in the sense that they are valid up to  $O(v^2/c^2)$, self energy effects contributing only to a 
higher order in this expansion parameter.

In the present paper we will address a particular case i.e. the stationary situation encountered 
in the rotation and 
revolution  around  the earth of an orbiting satellite (Lageos \cite{ciufo} and Gravity Probe B \cite{gravity}).

The same effects have also been accounted for \cite{veto1} \cite{veto2} within a gravitomagnetic (GEM) formulation of General Relativity (GR), confirming the GR  treatment \cite{schi} and thus making the curved spacetime picture seemingly unavoidable. 

We will comment on that and in particular on the cosmological treatment of the Coriolis effect.

On the contrary we will show that such an effect just comes from special relativity and that its {\it parameter free prediction } leads to an alternative simpler interpretation of physical reality.

As an extra outcome, also the gravitational clock effect will be given in a parameter free way.

\

\section{The vector equations}
 
The vector equations for gravitation are the following :
\begin{equation}\label{new}
\nabla\cdot\mathbf{g} = - 4 \pi  G  \rho
\end{equation}
\begin{equation}\label{car}
\nabla\cdot\mathbf{h} = 0
\end{equation}
in the first $\mathbf{g}$ representing the ``ordinary'' Newtonian field, while  the second for the gravitomagnetic 
field $\mathbf{h}$ based on the  \emph{assumption} (a fortiori even more reasonable than in electromagnetism) of the 
non existence of a gravitomagnetic charge.
 
These two are accompanied by the time dependent ones :
\begin{equation}\label{amp}
\nabla \times \mathbf{g} = -\frac{\partial\mathbf{h}}{\partial t}
\end{equation}
\begin{equation}\label{spo}
\nabla \times\mathbf{ h} = -\frac{4\pi G}{c^2}\,2\, \mathbf{j} + \frac{1}{c^2}\frac{\partial \mathbf{g}}{\partial t}
\end{equation}
The gravitomagnetic equation differs by the corresponding Maxwell one by the factor of $2$ in front of the ordinary 
mass current density  $\mathbf{j}$, required by special relativity.

Thus a post Newtonian formulation of gravitation has necessarily to embody a short distance repulsion from self energy 
effects (which modifies Newton's law) and velocity dependent, possibly repulsive terms, both effects, somewhat at 
variance with the standard picture, coming from elementary considerations. 

The time dependent terms which are crucial in determining the wave equation, play no role here, since we will consider 
only stationary conditions and the gravitomagnetic field generated by mass currents (self energy effects having been 
shown to be irrelevant), so that 
\begin{equation}\label{cur}
\nabla \times \mathbf{h} = -\frac{4 \pi G}{c^2}\,2\,\mathbf{j}
\end{equation}
thus implying a Lorentz gravitomagnetic force
\begin{equation}\label{lor}
\mathbf{F} =   m (\mathbf{g} + \mathbf{v} \times \mathbf{h})
\end{equation}
where $m$ is the relativistic mass. 

It is worth stressing that \emph{Eq.~(\ref{cur}) unambiguously determines the magnetic part of the Lorentz force 
Eq.~(\ref{lor})}, the product of $\mathbf{v}$ and $\mathbf{h}$ coming just from Lorentz transformations.  This point 
will be commented upon at length later on. 

Let us mention the extra constraint which additionally  backs up the present considerations. The induction law 
in its integral formulation, for the case of constant $\mathbf{h}$ and a varying circuit is in agreement with the 
Lorentz force \emph{only} in the present form. This represents therefore a double confirmation of the present formulas.

\section{Geodetic precession and frame dragging effects}

Let us then come to the Lageos \cite{ciufo} and Gravity Probe B experiments \cite{gravity}. As well known the latter
 measures the effects  of the orbital motion of the earth (of mass $M_{E}$) around the satellite (geodetic precession) and of its 
rotation (frame dragging) on satellite mounted gyroscopes at an altitude of  $642$ km. In both cases  the relevant 
parameter  which determines the angular velocities of the gyroscopes (apart from the numerical coefficients which will 
be given in the following) is, as usual,  
\begin{equation}\label{R_{S}}
\frac{GM_{E}}{c^2 R} \simeq 10^{-9}
\end{equation}
where $R \simeq 7000$ km. This, because of the preceding considerations about the successful effective vector formulation 
of gravity and the smallness of the effect, casts more than reasonable doubts as to whether these precessions should be 
unambiguously attributed to GR.

Thus one has for the gravitomagnetic field of a loop of radius $R$ described by the earth at the origin (i.e. the place 
of the satellite in its reference frame, around which the earth revolves) 
\begin{equation}\label{orb}
h_{orb} =  4 \frac{G \mu}{c^2  R^3} = 2\frac{G M_{E}}{c^2  R} \; \omega_{orb}
\end{equation}
with the  straightforward  dipole  extension to any direction.

The so called geodetic precession is simply due to the the angular velocity of precession of the (gravito) magnetic 
moment $\mathbf{\mu}$  of the satellite gyroscope (of standard angular momentum 
$\mathbf{S} = m r^2 \omega_{orb} \,\mathbf{n} =  2\mathbf{\mu}$) in a gravitomagnetic field $\mathbf{h}$ which is 
governed by the Newtonian equation 
\begin{equation}\label{torque}
\frac{1}{2}\mathbf{S} \times\mathbf{h} = \frac{\rmd\mathbf{S}}{\rmd t} 
\end{equation}
This implies 
\begin{equation}\label{om}
\Omega_{geo} = h/2
\end{equation}

To this \emph{spin orbit effect}, trivially governed by classical mechanism and SR (calculation of $\mathbf{h}$), one must 
add the Thomas (T) precession, again due solely to SR. An elementary derivation of the Thomas precession, in terms of proper 
time (the time on the satellite is not the time observed on the rotating earth which to a good approximation can be taken 
to be  that of the fixed stars) can be found in \cite{mash}, with the result 
$\Omega_{T} = -\frac{GM_{E}}{2R c^2}\omega_{orb}$.

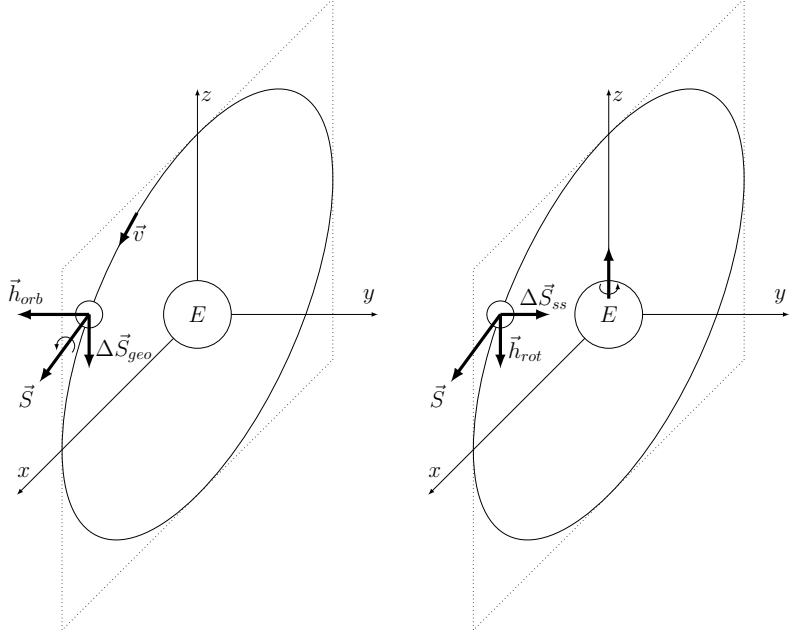
\begin{figure}[h]
\begin{center}
\scalebox{0.6}{
\begin{tikzpicture}[x=1.0cm, y=1.0cm]
\begin{scope}[>=latex]
\draw[->] (0,0)--(0,5);
\draw (0.2,4.85) node {\Large $z$};
\draw[->] (0,0)--(4,0);
\draw (3.8,0.4) node {\Large $y$};
\draw[->] (0,0)--(-4,-4);
\draw (-3.85,-3.5) node {\Large $x$};

\filldraw [fill=white] (0, 0) circle (0.75);
\draw (0, 0) node {\Large $E$};

\begin{scope}[yslant=1.0, xslant=0.0]
\draw (0,0) ellipse (3 and 4);
\draw [dotted] (3,-4)--(3,4)--(-3,4)--(-3,-4)--(3,-4);
\end{scope}

\filldraw [fill=white] (-2.4, 0) circle (0.3);

\draw [->, line width=2pt] (-1.34,2.25)--(-1.75,1.5);
\draw (-1.3,1.8) node {\Large $\vec{v}$};

\draw [->, line width=2pt] (-2.4,0)--(-3.5,-1.5);
\draw (-3.8,-1.8) node {\Large $\vec{S}$};
\draw (-3.12,-0.7) arc (180:-45: 0.2 and 0.2);
\draw [->] (-3.12,-0.7)--(-3.12,-0.8);

\draw [->, line width=2pt] (-2.4,0)--(-2.4,-1.2);
\draw (-2.38,-0.75) node[anchor=west] {\Large $\Delta\vec{S}_{geo}$};

\draw [->, line width=2pt] (-2.4,0)--(-4.,.0);
\draw (-3.8,0.1) node[anchor=south] {\Large $\vec{h}_{orb}$};
\end{scope}
\end{tikzpicture}
}
\hspace{0.2cm}
\scalebox{0.6}{
\begin{tikzpicture}[x=1.0cm, y=1.0cm]
\begin{scope}[>=latex]

\draw[->] (0,0)--(0,5);
\draw (0.2,4.85) node {\Large $z$};
\draw[->] (0,0)--(4,0);
\draw (3.8,0.4) node {\Large $y$};
\draw[->] (0,0)--(-4,-4);
\draw (-3.85,-3.5) node {\Large $x$};

\filldraw [fill=white] (0, 0) circle (0.75);
\draw (0, 0) node {\Large $E$};
\draw [->, line width=2pt] (0,0.35)--(0,1.5);
\draw (0.2,0.6) arc (0:-210: 0.2 and 0.15);
\draw [->] (0.2,0.6)--(0.2,0.7);

\begin{scope}[yslant=1.0, xslant=0.0]
\draw (0,0) ellipse (3 and 4);
\draw [dotted] (3,-4)--(3,4)--(-3,4)--(-3,-4)--(3,-4);
\end{scope}

\filldraw [fill=white] (-2.4, 0) circle (0.3);

\draw [->, line width=2pt] (-2.4,0)--(-3.5,-1.5);
\draw (-3.8,-1.8) node {\Large $\vec{S}$};

\draw [->, line width=2pt] (-2.4,0)--(-2.4,-1.2);
\draw (-2.35,-0.75) node[anchor=west] {\Large $\vec{h}_{rot}$};

\draw [->, line width=2pt] (-2.4,0)--(-1.3,0);
\draw (-2.1,0.0) node[anchor=south west] {\Large $\Delta\vec{S}_{ss}$};
\end{scope}
\end{tikzpicture}
}
\end{center}
\caption{\emph(Left) The satellite spin $\mathbf{S}$ lies in the $(x,z)$ plane described by its orbital motion around the 
earth. The gravitomagnetic loop described by the latter around $\mathbf{S}$, generates on it $h_{orb}$ perpendicular 
to the plane which makes the gyroscope to precess, even with the inclusion of the Thomas recession.\newline 
\emph{(Right)} The spinning earth generates an additional gyromagnetic field $h_{rot}$ on the spinning gyroscope 
$\mathbf{S}$, this time in the orbital plane. The gravitational spin-spin interaction makes the plane rotate around $z$ 
(frame dragging). Both effects are predicted by SR in a flat Euclidean space}
\end{figure}

Notice that since the satellite is seen to precess, in its reference frame it recedes with respect to the fixed stars. 
Thus the total angular velocity of precession due to the earth revolution is
\begin{equation}
\frac{\Omega_{geo}}{\omega_{orb}}=\frac{1}{2} \frac{GM_{E}}{c^2 R}  
\end{equation}

i.e. a relative effect determined by the (weak) gravitational field strength parameter, as illustrated in Fig.(1) Left. 

The present result Eq.~(\ref{om}) differs by a factor of $2$ from the GR calculations by Schiff \cite{schi}.   

Also numerous NR reductions of GR for the weak field and low velocity case have been recently appeared 
\cite {lano, bed, kol, mas, ago}. Apart from their problems with wave propagation, critically commented upon in
\cite{christbara}, they seem to confirm Schiff's result only by introducing an extra factor of $2$ in the Lorentz 
force Eq.~(\ref{lor}).

As underlined before, this is forbidden just by SR transformations which connect Eq.s~(\ref{new}), ({\ref{cur}}), 
({\ref{lor}}), and ({\ref{amp}).

Let us then consider the smaller effect due to the magnetic field created by the earth rotation around its axis: the 
\emph{spin-spin effect} usually dubbed ``frame dragging effect''.
It goes without saying how special relativity is again all one needs.
The gravitomagnetic dipole moment of a mass element $\rmd m$ of a rotating body is  
$\rmd \mu =\frac{1}{2} \omega r^2 \rmd m$, so that the gravitomagnetic field of the spinning earth at the gyroscope
in an arbitrary direction reads 
\begin{equation}
\mathbf{h}_{rot} =\frac{G I_{E}}{c^2 R^{3}} \, [(3\mathbf{\omega}_{rot}\cdot\mathbf{n})\mathbf{n} - \mathbf{\omega}_{rot}] 
\end{equation}

$\mathbf{n}$ standing for the unit vector along  $\mathbf{R}$.
Now the previous torque equation (\ref{torque}) obtains so that one immediately gets the resulting precession in terms of
the angular velocity of rotation of the earth $\omega_{rot}$ 
\begin{equation}
\Omega_{spin-spin } = h_{rot} /2
\end{equation}
Its direction, at right angles with the geodetic precession, is shown in Fig. (1) Right.  
This time again a direct (i.e. without the intermediary of the Thomas precession) factor of $1/2$ results from the 
comparison with the existing literature.

\section{Discussion}
  
As mentioned our predictions differ from the quoted measurements by a factor of $1/2$.  
Some extensive comments are then in order. 
  
The GR based predictions have been confirmed to a different degree of accuracy by the Lageos \cite{ciufo} and Gravity 
Probe B experiments \cite{gravity} and will be further scrutinized by the proposed multi-ring-laser underground
experiment \cite{pisa}. 

The situation appears hence somewhat contradictory.  NR reductions of GR equations give a vector formulation which 
(while confirming the soundness of the present approach) is in agreement with GR Schiff's results only at the price 
of a wrong Lorentz force ! 

Therefore, granting the correctness of the experimental results and apart from it, the basic question we have 
to address is: is the doubly rotatory motion of the satellite gyroscope $S$ determined only by the earth motion ?

It is indeed clear that the gyroscopes are just an up to date version of Newton's bucket.  Therefore, if in line 
with Mach's thinking, we do not believe in absolute motion we have to ascertain the role of the relative motion of 
rest of the Universe. This time quantitatively, since the presumed sole (and dominant) contribution of the earth has 
a quantitative estimate. 

The point is that the free fall satellite frame is an inertial one so long as it does not rotate. Once it does, due to 
the earth effect, it no longer is. We must therefore introduce Coriolis forces or the effect of the rest of the Universe.
\begin{equation}
\mathbf{F}_{Cor}  =   2 m \, \mathbf{v} \times\mathbf{\omega}
\end{equation}
from which \cite{ahah}
\begin{equation}
\mathbf{ M}_{Cor} = \mathbf{S} \times\mathbf{\omega}
\end{equation}
where $\omega$ refers respectively (and separately) to each of the two rotations induced by the movement of the earth. 
Thus we have to add to  Eq.~(\ref{torque})  this extra contribution,  obtaining  a total rotation 
\begin{equation}
\Omega  = h
\end{equation}
instead of the previous  $\Omega= h/2$.

Thus our prediction of the spin spin precession is simply doubled whereas for the geodetic precession the doubling of 
$\Omega$ combined with the unaffected Thomas precession yields a final factor of $3/2$ for $\Omega_{geo}$, this time 
again in accord  with Schiff's result.

The previous result provides a deeper understanding of the (non) equivalence principle: \emph{the fact that 
forces are locally eliminated in the free falling frame  (no tide effects), does not imply the same for the moments !}
\cite{cor}

We are then led to revisit Sciama's conjecture \cite{sciama}  who has greatly emphasized the similarity between the 
previous gravitomagnetic force and the ``fictitious'' Coriolis force experienced in a rotating frame, 
stressing \emph{the connection between angular velocity of rotation and corresponding magnetic field. 
The proportionality coefficient being simply given by the ubiquitous factor $GM/(c^2 R)$!}

Indeed as an extension to the (rest of the) Universe of the  previous expression for the gravitomagnetic field of
a mass $m$ it  follows
\begin{equation}\label{bri}
\mathbf{F}_{GM}  =    m\, \mathbf{v} \times\left( \frac{2 G M}{c^2 R}\mathbf{\omega}\right) = 2 mv \frac{GM}{c^2R} 
\omega \, \mathbf{n}
\end{equation}
the suffix GM standing for gravitomagnetic.
 
Thus if 
\begin{equation}\label{U}
\frac{GM_{U}}{c^2 R_{U}} =1 
\end{equation}
then
\begin{equation}
\mathbf{F}_{Cor}  =   \mathbf{F}_{GM} 
\end{equation}
The essential point in this argument is that in the relative rotation of the satellite with respect to the Universe, the 
magnetic field generated by distant layers of matter goes as $1/R$ i.e. the same behaviour of radiation, rather than the 
usual $1/R^2$ of Newton  forces.  Therefore a relative more important role even of distant stars is a matter of fact.  
 
In favour of the estimate/Ansatz of Eq.~(\ref{U}) there is a lot of circumstantial evidence as well as speculations 
\cite{christI} \cite{christII}. In particular it is necessary to account for the precession of the Foucault pendulum as determined 
along the present lines by the rotating matter of the Universe.

\section{The gravitational clock effect}
 
With inclusion of the gravitomagnetic force, the two body gravitational equation of motion thus reads
\begin{equation}\label{gmag}
m \omega^2 R = \frac{GMm}{R^2} + m v h 
\end{equation}
where the sign of the last term, depending on the relative orientation of the velocity of the mass $m$ orbiting 
the spinning mass $M$ will be detailed at the end. For the case of  satellites in the equatorial 
plane $h = h_{rot}$ 
is given by\begin{equation} \label{hrot}
\mathbf{h}_{rot} = - \frac{G I_{E}}{c^2 R^{3}} \,\mathbf{\omega}_{rot}
\end{equation}
Here the Thomas precession which affects in the same way both satellites has been omitted. This academic case, upon which we will comment later, is considered just for comparison with the existing literature \cite {clock}.

Thus in terms of the angular momentum $S$ of the spinning mass $M$ (the earth), of the Keplerian angular velocity 
$\omega_{K} = \sqrt {GM/R^3}$ and of the post Keplerian correction $\tau$    
\begin{equation}
\tau = S/ M c^2 
\end{equation}
one has
\begin{equation}
\omega^2  = \omega_{K}^2+ \omega_{K}^2 \omega\; \tau
\end{equation}
 
The admissible root is 
\begin{equation} \label{delom}
\omega  \simeq \omega_{K} + \frac{1}{2}\omega_{K}^2 \, \tau
\end{equation} 
where terms of higher order in $\tau$  have been neglected. 
 
Thus
\begin{equation}
T = \frac{2 \pi}{\omega}  =\frac{2 \pi}{\omega_{K} +\frac{1}{2}\omega_{K}^2 \,\tau} \simeq  T_{K}  - 2 \pi\frac{\tau}{2}
\end{equation} 
 
As in the preceding case the result differs by the GR prediction  by a factor of $1/2$.
Some comments are  therefore in order. Eq.(\ref {gmag}) holds true in an inertial reference frame i.e. for the so called fixed stars, so that what might in principle be observed on earth has to be corrected, as usual,  for its (complicated) motion.

The point is thus whether the magnetic effects of its rotation $ O(10^{-9})$ have been taken into account or not. 

Notice that the correction  term to the Keplerian angular velocity in Eq.( \ref {delom}) is closely related to the gravitomagnetic field

\begin{equation}
\omega' = \frac{1}{2}\omega_{K}^2 \, \tau = \frac{G S}{2 c^2 R^{3}} = \frac {{h}_{rot} }{2}
\end{equation} 
(in a first approximation we do not distinguish between the orbiting satellite radius R and the earth radius $R_{E}$).
 
Now do the corrections of the earth motion with respect to the fixed stars include or not its gravitomagnetic effects ? It seems plausible, apart from practical problems, that not to be the case also as a matter of principle. 

Indeed in the presence of a third body (the fixed stars) the "free fall" of the orbiting satellite is not really such.
The velocity of all point-like bodies would be indeed the same but it would be different even along the same orbit in the reversed direction, thus questioning the validity of the use even locally of  inertial frames.  

This is an extra  illustration, in addition to the one given in connection with the gyroscopes, of how, to a closer scrutiny, the principle of equivalence be just a very good first order  approximation and of how, like the other forces, gravity cannot be eliminated. 

In this case one should add a "gravitomagnetic" Coriolis term so that the total equation of motion would read

In other words the complete projected equation of motion reads 
\begin{equation}
 \omega^2 =\omega_{K} ^2 +  \omega h + 2 \omega \; \omega'
\end{equation}
which would thus double the previous contribution yielding 
a post Keplerian correction 
\begin{equation}
\Delta T _{p K} = \pm \; 2  \pi  \tau 
\end{equation} 
which agrees  with the GR result \cite {clock}.
 
The plus sign applies for the same sense of rotation of the satellite and the earth, whereas the minus 
(smaller period) for antirotation.
 
We are therefore in the presence, in principle, of an additional test of gravitomagnetism where SR is enough to 
predict the results of GR.

Coming to facts, in addition to the many effects (thoroughly considered in \cite {clock}) which might affect an experimental test,  the possibility of synchronizing two satellites seems rather remote. One should therefore  use just one and in this case the Thomas (T) precession should be considered.
This would yield a correction 
 \begin{equation}
\Delta T _{T} = T_{K} \times \frac {GM}{2 c^2 R} 
\end{equation} 
which definitely competes with the previous post Keplerian correction. 
 
\section{Conclusions }

Gravitomagnetism resulting from SR yields a set of parameter free vector equations which provide an effective theory 
of gravitation.

They have been shown to predict in elementary terms the quadrupole gravitational radiation in a flat Minkowski space. 

In this work the particular case of stationary currents has been considered and applied to geodetic precession, frame 
dragging and the gravitational clock effect.  

It has been shown, contrary to naive expectations, that the orbital and spin rotation of the earth do not account for 
the experimental results, the contribution of the rotating Universe on the non inertial frame of the satellite being 
of the same order of magnitude of the earth's.

The following comments are inevitable :
\begin{itemize} 
\item in this case SR is all we need to get the GR results
\item the Machian picture gets a piece of support and the role of counterrotating fixed stars is paramount
\item GR, in spite  of its claims of generality, assumes a privileged reference frame \cite{mach} ! Empirically, this system 
coincides with the average system of the fixed stars, however, this correspondence appears incidentally, since 
the presence of the distant masses did in no way enter the calculation.  
\end{itemize}

\emph{Thus Mach's thinking enters quite rightly and quantitatively our picture of the Universe through the prediction 
of the ``fictitious''  Coriolis force in a post Newtonian language which, in our opinion, has the advantage, besides 
its simpler formulation, of  making such a connection plain ! }
 
Therefore it is really rewarding to have such a deep link between local and global properties of the 
Universe.

We must unescapably accept the existence of a privileged frame of reference: the microwave background radiation, to a 
very good degree of accuracy, takes the place and confirms the ``fixed stars system''! 

\
 
\hspace{0.5cm} 
ACKNOWLEDGMENTS
  
It is a pleasure  to thank G.~Morchio,  G.~Cicogna  and C. Bonati for their interest and help in this work and C. Bonati for 
assistance  with the figures.

\


\begin{thebibliography}{999}

\bibitem{christbara}
P.~Christillin,  L.~Barattini arXiv:1205.3514v1 (2012)

\bibitem{ciufo}
I.~Ciufolini The 1995-99 measurements of the Lense- Thirring effect using laser-ranged satellites, Class. Quan- tum Grav. 17 2369 (2000) 
 
\bibitem{gravity}
C.~W.~F.~Everitt et al. PRL {\bf 106} , 221101 (2011)


\bibitem{veto1}
B. Veto" Eur.J.Phys. 31 (2010) 1123-1130

\bibitem{veto2}
B. Veto" Eur.J.Phys. 32 (2011) 1323-1329

\bibitem{schi}
L.~I.~Schiff 1960 Proc. Natl Acad. Sci. 46 871Ð82

\bibitem{mash}
G.~F.~Smoot, http://aether.lbl.gov/www/classes/p139/homework/seven.pdf



\bibitem{pisa}
F.~Bosi et al. Phys.Rev D 84 122002 (2011)



\bibitem{MTW}
C.~W.~Misner, K.~S.~Thorne and  J.~A.~Wheeler,  Gravitation, W.H. Freeman and Company, San Francisco 1970

\bibitem{lano}
R.~P.~Lano (1996). "Gravitational Meissner Effect". arXiv:hep-th/9603077

\bibitem{bed}
D.~Bedford and P.~Krumm, Am. J. Phys. 53, 889, (1985)

\bibitem{kol}
H.~Kolbenstvedt,  Am. J. Phys. 56, 523 (1988)

\bibitem{mas}
B.~Mashhoon et al. 2000 arXiv:gr-qc/991202

\bibitem{ago}
M.~Agop et al. arXiv:gr-qc/9911011

\bibitem{clock}
B.~Mashhoon et al. 1998 arXiv:gr-qc/9008017
 
\bibitem{cor}
This effect has not been considered so far.  Indeed Schiff explicitly writes that the satellite is ``in free fall''. The 
precession takes place with respect to the inertial frame, which is generally believed to be defined by the distant 
extragalactic nebulae, the so called ``fixed stars''. 

\bibitem{ahah}
This apparently contradictory result is due to the elementary but sometimes overlooked fact that whereas 
$\mathbf{v} =\mathbf{\omega} \times \mathbf{r}$, $\mathbf{\omega} =\frac{1}{2}\mathbf{r}\times\mathbf{v} $ .

\bibitem{sciama}
D.~W.~Sciama, The unity of the Universe,faber and Faber, London, 1959

\bibitem{christI}
P.Christillin,  EPJ Plus (2011) {\bf 126} 48


\bibitem{christII}
P. Christillin,  EPJ Plus (2011) {\bf 126} 88


\bibitem{mach}
H.~Lichtenegger and B.~Mashoon, 
arXiv:physics/0407078 [physics.hist-ph]  ``Ironically, though general relativity was intended to be based on relational 
concepts, contrary to its name it still contains absolute elements . This is already expressed in the calculation of the 
advance of Mercury's perihelion, which is referred to a coordinate system ..''

\end{thebibliography}
\end{document}


\title{Addendum to " The Machian contribution of the Universe to .."
\\
The Machian origin of the centrifugal force }

\author {\\ \\
L. Barattini\\
Universit\`a di Pisa\\
\\and\\
\\P. Christillin\\ \
Dipartimento di Fisica, \\
Universit\`a di Pisa\\
I.N.F.N. Sezione di Pisa\\
\\ \\}

\maketitle

\begin{abstract}
A  derivation of the centrifugal force from an effective vector formulation of gravitation is attempted. The centrifugal force appears to be due  to a relativistic effect of the counter-rotating Universe.     Gravitomagnetic energy effects, a million times stronger than the self energy effects responsible for curvature in the GR language, would thus produce the centrifugal acceleration.   The Machian picture, already successful in the case of  the Coriolis force,  gets an additional circumstantial support. 
\end{abstract}

\

\

\newpage

\section{Introduction}

\

In spite of centennial speculations \cite{Berkeley}  \cite{mach},  a satisfactory, at least semiquantitative, solution of the problem of relative rotational motion is, in our opinion, still lacking.

Is the rotation of , say, the earth with respect to the rest of the Universe equivalent to a counter rotation of the latter ?

Further arguments in favour of this logically stringent position have been put forward more recently by Sciama \cite{sciama}, who however has not gone farther than stressing the analogy of gravitomagnetic with magnetic forces, thus making {\it plausible} such an origin in the case of the Coriolis force. 

This has been proven by us in \cite{christbara2}.

General relativity (GR) does not address the problem at all, since  in its privileged frame of reference (" the 
precession takes place with respect to the inertial frame, which is generally believed to be defined by the distant 
extragalactic nebulae, the so called " fixed stars''  \cite{schi}) no mention is made of the rest of the Universe.

According to  \cite{mash} also " Ironically, though GR was intended to be based on relational 
concepts, contrary to its name it still contains absolute elements . This is already expressed in the calculation of the 
advance of Mercury's perihelion, which is referred to a coordinate system ..''  

The aim of the present paper is to extend the considerations already used in \cite{christbara2} to account for  the Coriolis force, to predict unavoidably the{ \it form} of the centrifugal force and to show that its coefficient is, within the present Universe estimates, compatible with the canonical value.

The essential points will be:

i) the proportionality between the gravitomagnetic field of a rotating mass distribution and its angular velocity and their dimensional equivalence 

ii) the expression of the gravitomagnetic energy density

iii) the kinematical relation among quantities in inertial and rotating frames by which the centrifugal acceleration  can be linked to the gravitomagnetic field (our final equation).

\

\section{The centrifugal force from the counter rotating Universe}

\

In two recent works of ours \cite{christbara}, \cite{christbara2} 
 a  set of  effective vector equations for low velocity weak field gravitation  has been \emph{derived} from special relativity and 
shown to predict in simple terms the quadrupole gravitational radiation as well as  geodetic precession, frame dragging and the gravitational clock effect. 

Numerous NR reductions of GR for thesame conditions have been recently appeared 
\cite {lano, bed, kol, mas, ago} confirming  the  soundness of such an approach.

Most important, in respect to the matter we are addressing here, the Coriolis force (since the equivalence principle is explicitly used  we will speak indifferently of force and acceleration) has been shown to play a crucial role  in the above mentioned stationary processes and the role of the (rest of the) Universe to be crucial in explaining the observed effects.

Indeed the gravitomagnetic (GM) force of a rotating mass $M$, at a distance $R$,   on a test mass $ m$ reads
\begin{equation}\label{bri}
\mathbf{F}_{GM}  =    m\, \mathbf{v} \times\left( \frac{2 G M}{c^2 R}\mathbf{\omega}\right) = 2 m\, \mathbf{v} \times\left(\frac {G M}{c^2 R}\right) \mathbf{\omega}
\end{equation}

Indeed the gravitomagnetic (GM) force of a rotating mass $M$, at a distance $R$,   on a test mass $ m$ reads
\begin{equation}
\mathbf{F}_{Cor}  =   2 m \, \mathbf{v} \times\mathbf{\omega}
\end{equation}

Thus, when applied to the Universe, if 
\begin{equation}\label{U}
\frac{GM_{U}}{c^2 R_{U}} =1 
\end{equation}

and this relation compares favourably with present day estimates as well as with other theoretical considerations \cite{christI} \cite{christII} , it follows that 

\begin{equation}
\mathbf{F}_{Cor}  =   \mathbf{F}_{GM} 
\end{equation}
The relevant  point in this argument is that in the relative rotation,  the 
magnetic field generated by distant layers of matter goes as $1/R$ i.e. the same behaviour of radiation, rather than the 
usual $1/R^2$ of Newtonian  forces.  Therefore a relative more important role even of distant stars is a matter of fact.  

Thus the physical origin of the Coriolis force seems to get  a semiquantitative confirmation.

Let us pass over to the centrifugal force with some additional remarks. 

Now whereas  a gravitomagnetic origin of a Coriolis force might seem reasonable (effect of counterrotating masses on a moving one), at first sight  it might seem puzzling the  effect of the same  counterrotating masses on a mass  in its rest frame. As it has been pedagogically underlined  in  \cite{christbara} however a mass at rest experiences a force from the relativistic effects (i.e. $O(v^2/c^2)$ ) of moving ones (even if this is customarily expressed as magnetic force). 
And indeed the relativistic origin of the effect is evident from the
  proportionality coefficient  $\frac{GM}{c^2 R}$!

The essential point in the previous considerations  is that a rotating matter distribution produces a gravitomagnetic field $\mathbf{h}$ proportional to the angular velocity of rotation $\mathbf{\omega}$

\begin{equation}\label{U}
\mathbf{h} \propto \mathbf{\omega}
\end{equation}

the proportionality coefficient depending of course on the geometry (loop, spherical shell, etc.).  In other words a gravitomagnetic field produced by moving masses is dimensionally equivalent to an angular velocity. 

This has a profound physical meaning.  We know that the $ T \neq 0$ cosmic background radiation, essentially coincident with the fixed stars system, represents {\it the} privileged inertial reference frame. However in terms of relative motion  the fact that the rotation of the Universe, as seen from us, be determined by the properties of the other masses ($M_{U}$ and $R_{U}$) renders physical what seemed just a kinematical affair. 

Therefore if the previous relation between $\mathbf{ h}$ and $ \mathbf{\omega}$ holds true, just a two-fold application of the kinematical relation for operators
 
 \begin{equation}
\left(\frac{d}{dt}_{(I)}\right) =  (\mathbf{\omega}  \times)  + \left(\frac{d}{dt}_{(R)}\right)
\end{equation}
(where the suffixes refer respectively to the inertial (I) and rotating (R) frames) yields for the acceleration of the radius vector $\mathbf{ r}$ the additional centrifugal acceleration.

Let us give some additional arguments. 

Consider a symmetric spherical rotating shell.
Its mass (energy and mass are used indifferently) density reads \cite{christbara}

\begin{equation}
\rho_{h}  =  -  \frac{1} {\;4 \pi G}\: \frac{h^2} {2}
\end{equation}

The Coriolis force has been accounted for by a gravitomagnetic field where for the contribution  of the Universe the same expression obtained for a mass loop (the orbiting earth) where  $ \mathbf h = \dfrac{2\; GM}{\; c^2R}\: \mathbf{\omega}$ has been used.  On the contrary if  one considers spherical symmetry,  in the interior a {\it constant} gravitomagnetic field (see e.g. \cite{grif})
 $\mathbf h = \dfrac{4\; GM}{3\; c^2R}\: \mathbf{\omega}$.  If we use the  value 2 which reproduces the Coriolis force, then 
from the expression of the field energy ($ U = \frac{4 \pi   r^3 \rho_{h}}{ 3 r} $) one  gets  

\begin{equation}
F_{C}   =  2/3 \; m \;
\mathbf{\omega}  \times ( \mathbf{\omega}  \times \mathbf{r} )
\end{equation}

a {\bf centrifugal} force due to the {\bf negative} energy density.
This result is noteworthy in many respects.

  First {\bf the centrifugal force is a relativistic effect ! }

Second,  the correct dimensional requirement for the acceleration comes from a (subtle ?) interplay between the expression for the mass density and that for the field, which makes the desired $\omega^2$ factor unavoidable. Moreover  the gravitational constant $G$ only enters through the standard weak field formula in brackets. 
The coefficients, upon whose evaluation many criticisms might apply, is remarkably close to one.

In this respect let us once more underline how even two drastically different density expressions like $ \rho \simeq constant $ and  $ \rho \simeq 1/r^2$, which implements the black hole possibility,  yield for the self energy the two very close coefficients 3/5 and 1 respectively. Thus even if our evaluation of the total Universe contribution by simply substituting its values is surely questionable,  the semiquantitative agreement can hardly be regarded as fortuitous. 

Thus the fact that one cannot reproduce at the same time, within the same approximations, both the coefficients of the Coriolis and of the centrifugal force appears less unsatisfactory.  Also the comparison with the Lense- Thirring \cite {lense} work (where a factor 4/15 appears. See \cite {pfi} for comments) backs up the previous statement.

The reason why only gravitomagnetic forces act is obvious : within a symmetric spherical shell the static gravitoelectric effects cancel out because of the symmetry, whereas the magnetic ones, constant in $R$, are different from zero and along $\mathbf{\omega}$.  

The fact that no retardation for magnetic terms is present, depends on our choice of the gauge, as explained in \cite{christbara}, see also  \cite{fey}.

\
\

\

\

\begin{figure}[!htbp]
\centering
\scalebox{0.2}{\hspace{-18cm}\includegraphics{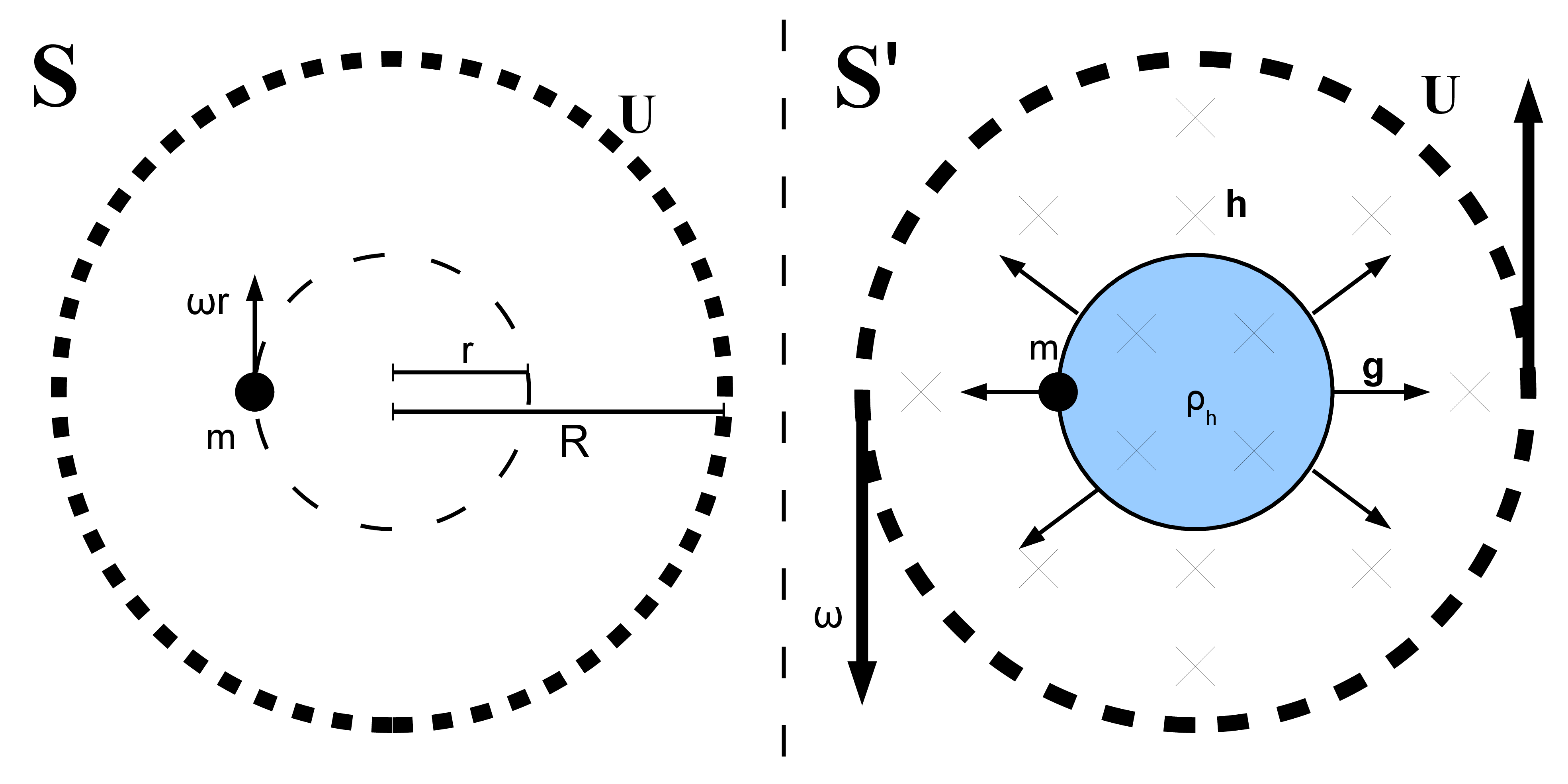}}
\caption{
A mass $m$ rotates in the fixed Universe frame $S$ at a distance $r$ from the center.  From the mass rest frame $S'$ the Universe is seen to counterrotate, generating a gravitomagnetic field $\mathbf{h} \propto \mathbf{\omega}$
and a gravitomagnetic field density which causes the "fictitious" centrifugal force.  As a particular case our mass is at the surface of the earth and the whole Universe contributes to the repulsion. Thus for  the Earth one has the fascinating fact of a gravitoelectric self energy (space curvature) effect of $O(10^{-9})$ with respect to $g$, and of a much bigger gravitomagnetic influence , due  to its rotation (or better to the counterrotation of the Universe), of  $O(10^{-3})$. Also in the former case self energy acts "centrifugally"  so as to diminish $g$.
}
\label{fig:Centrif}

\end{figure}
\

\

\

\

\

\

\section{Conclusions}

The fact that only {\it relative rotations} have a physical significance has thus been substantiated, both as regards the expression of the centrifugal force as well as its actual value. 

Some more comments are in order. 

It is not superfluous to underline the similarities and  differences with the case of orbiting satellites \cite{christbara2}. There for the gyroscopes in free fall around the earth the effect of the Universe rotation provided only part of the effect (essentially 1/2) the other being due to the earth rotation.  Here of course only the former contributes both for moving objects (Coriolis on the earth) and for masses at rest (centrifugal).  Thus this double constraint gives us some more confidence in a non accidental agreement.

Therefore it is really rewarding to have such an interesting  link between local and global properties of the 
Universe and probably a deeper understanding of gravitoelectric effects (self energy or space time curvature where only the earth constituents are involved) and of the gravitomagnetic  ones (much bigger centrifugal acceleration determined by the Universe).

\emph{ In conclusion Berkeley -Mach's \cite{Berkeley} \cite{mach}
thinking enters quite rightly our picture of the Universe through the prediction, in addition to the Coriolis,  also of the 
 ``fictitious''  centrifugal force as "real ones" !}

\

\

ACKNOWLEDGMENTS

It is a pleasure to thank C. Bonati, G. Cicogna and G. Morchio for a critical reading of the manuscript and for helpful comments.

\